\documentclass[preprint2]{aastex6}

\usepackage{graphics}
\usepackage{natbib}
\usepackage{amsmath}

\shortauthors{Tamayo et al.}

\begin{document}

\title{A Machine Learns to Predict the Stability of Tightly Packed Planetary Systems}

\author{Daniel Tamayo\altaffilmark{1,2,3}, Ari Silburt\altaffilmark{1,4}, Diana Valencia\altaffilmark{1,4}, Kristen Menou\altaffilmark{1,4}, Mohamad Ali-Dib\altaffilmark{1,2,3}, Cristobal Petrovich\altaffilmark{2,5}, Chelsea X. Huang\altaffilmark{1,3,4}, Hanno Rein\altaffilmark{1,4}, Christa van Laerhoven\altaffilmark{2,5}, Adiv Paradise\altaffilmark{1,4}, Alysa Obertas\altaffilmark{4} and Norman Murray\altaffilmark{2}}
\altaffiltext{1}{Department of Physical \& Environmental Sciences, University of Toronto at Scarborough, Toronto, Ontario M1C 1A4, Canada}
\altaffiltext{2}{Canadian Institute for Theoretical Astrophysics, 60 St. George St, University of Toronto, Toronto, Ontario M5S 3H8, Canada}
\altaffiltext{3}{Centre for Planetary Sciences Fellow}
\altaffiltext{4}{Department of Astronomy \& Astrophysics, University of Toronto, Toronto, Ontario M5S 3H4, Canada}
\altaffiltext{5}{CITA Fellow}
\email{d.tamayo@utoronto.ca}

\begin{abstract}
The requirement that planetary systems be dynamically stable is often used to vet new discoveries or set limits on unconstrained masses or orbital elements. 
This is typically carried out via computationally expensive N-body simulations.
We show that characterizing the complicated and multi-dimensional stability boundary of tightly packed systems is amenable to machine learning methods. 
We find that training an XGBoost machine learning algorithm on physically motivated features yields an accurate classifier of stability in packed systems. 
On the stability timescale investigated ($10^7$ orbits), it is 3 orders of magnitude faster than direct N-body simulations. 
Optimized machine learning classifiers for dynamical stability may thus prove useful across the discipline, e.g., to characterize the exoplanet sample discovered by the upcoming Transiting Exoplanet Survey Satellite (TESS).
This proof of concept motivates investing computational resources to train algorithms capable of predicting stability over longer timescales and over broader regions of phase space.
\end{abstract}


\section{Introduction} \label{intro}
In order to characterize planetary systems, it is common practice to assume long-term stability in order to set upper limits on planetary masses and orbital eccentricities \citep[e.g.][]{Lissauer11, Steffen13, Tamayo14b, Tamayo15}.
This involves running grids of direct N-body integrations over the large multi-dimensional parameter space of initial conditions that are consistent with observational error.
In practice, however, one can often explore only a minute fraction of the phase space; in the case of many systems discovered by the Kepler mission, each integration requires several weeks of computation to simulate timescales comparable to the star's age ($\gtrsim 10^{11}$ planetary orbits) with current hardware.
A more efficient classifier of dynamical stability would thus be invaluable.

In this investigation, we specialize to the case of tightly packed systems, where the interplanetary separations are less than 10 mutual Hill radii ($R_H$), where
\begin{equation}
R_H \equiv \left(\frac{M_1+M_2}{3M_\star} \right)^{1/3} a_1,
\end{equation}
$M_1, M_2$ and $M_\star$ are the masses of a pair of planets and the central star, and $a_1$ is the semimajor axis of the inner planet.
This regime has long been recognized as important in the early stages of planetary systems when bodies are still merging, and has received much attention.

For the special case of two-planet systems, there exists an analytic ``Hill criterion" that, if satisfied, precludes close encounters between the planets for all time (\citealt{Marchal82}, \citealt{Milani83}, \citealt{Gladman93}, but see also \citealt{Barnes06}, \citealt{Deck12} and \citealt{Veras13hill}).
However, in the general case with more than two planets, the additional degrees of freedom preclude a topological criterion for Hill stability.

This has led many authors to perform suites of N-body simulations and fit empirical curves to the results.
Because of the prohibitively large phase space of possible initial conditions, several authors \citep[][Obertas et al., submitted]{Chambers96, Faber07, Smith09} have considered the special case of initially circular, planar orbits with all planet pairs having equal Hill separations $\Delta$, defined as the difference in semimajor axis divided by $R_H$.
They find that instability timescales grow approximately exponentially with Hill separation; however, the fitted coefficients change as the number of planets and planetary masses are varied, and introducing inclinations \citep{Marzari02}, eccentricities \citep{Ito99, Chatterjee08, Pu15}, or unequal spacings between planets \citep{Marzari14} change the instability timescales substantially.
For a given planetary system, it is therefore not always clear which scaling law is appropriate to apply, and what confidence one can have in the resulting estimate.

For this investigation we take a new machine learning approach.
High dimensional classification tasks of this kind are ubiquitous across industry and data science, and sophisticated machine learning algorithms have been developed to tackle these problems.
Such techniques have been highly successful in an astronomical context for several image classification tasks, e.g., assigning morphological types to galaxies \citep{Collister04}; however, they have seen little use in dynamical classification to date\citep[see][for a recent counterexample]{Petrovich15}.

\section{Methods}
We choose to frame the problem as a binary classification task, i.e., predicting whether or not a given planetary system is stable (over a given timescale).
Each ``example" (planetary system) is described by a set of ``features" that the algorithm uses to predict stability, in the form of a probability between 0 and 1.
In supervised machine learning, an algorithm is first trained on examples where it is told the correct answer (stable or not stable).
The trained algorithm can then be used to predict on new examples.

\subsection{Dataset} \label{dataset}
In order to train our algorithms, we generated a dataset of 5000 N-body integrations of 3-planet systems over $10^7$ orbits of the innermost body.
We focus on 3-planet systems since there exists an analytic criterion for the case of two planets \citep{Gladman93}, and systems with more planets exhibit qualitatively similar behavior \citep{Chambers96}.
The number of simulations and length of integration were chosen to generate a dataset at limited computational cost ($\sim 1000$ CPU hours), and assess the value of investing significant computing time to train classifiers on more astrophysically relevant timescales ($\sim 10^9$ orbits).
Because we expect that instability timescales of $10^7$ and $10^9$ orbits are both physically driven by Chirikov diffusion due to the overlap of mean-motion resonances (see Fig.\:2 in Obertas et al., submitted), we expect the performance of models trained on this dataset to be comparable to that of similar algorithms trained on longer datasets.

With a view toward applying these models to Kepler discoveries, we adopted a solar-mass star, 5 $M_\Earth$ (Earth-mass) planets, and drew the innermost planet's semimajor axis randomly between 0.04 and 0.06 AU.
We note, however, that our results are strictly scale-free, and can be applied to comparable systems with masses, orbital periods and semimajor axes expressed in terms of the star's mass, innermost planet's orbital period and semimajor axis, respectively.
The second planet's semimajor axis was separated from the first by a number of mutual Hill radii drawn from a uniform distribution in the range $[5,9]$.  
The third planet's separation was then independently drawn from the same distribution, yielding unevenly spaced planets.
The particular range of Hill-radius separations was chosen to capture the regime of interest on our adopted timescale of $10^7$ orbits and roughly generate a balanced dataset of stable and unstable systems (this yielded 1479 stable systems out of 5000).
Eccentricities and inclinations were drawn independently for each planet from uniform distributions over [0, 0.02] and [0, 1$^\circ$], and the remaining angles were drawn randomly over [0,$2\pi$].

Because one only needs to integrate until the first Hill sphere crossing (once this happens, strong scatterings happen quickly, e.g., \citealt{Gladman93}), we use the efficient {\sc \tt WHFAST} integrator \citep{ReinTamayo15} in the open-source {\sc \tt REBOUND} N-body package \citep{Rein12}. 
{\sc \tt REBOUND} is written in \texttt{C99} and comes with an optional \texttt{python} interface.
We adopted a conservative timestep of 1\% of the innermost planet's orbital period, and classified systems as unstable if any pair of planets came within 1 Hill radius of each other during the simulation.

\subsection{Metrics} \label{metrics}
Binary classifiers are often evaluated on their {\it precision} (here the fraction of systems that are actually stable when the model predicts stability) and {\it recall} (the fraction of systems the model predicts are stable out of the truly stable cases).  
For typical algorithms that predict probabilities of class membership, one can trade off between precision and recall by varying the threshold probability for classification.
For example, a conservative model that only classifies systems as stable if it predicts a probability of stability greater than 0.99 will be right most of the times that it predicts stability (high precision), but will miss all the stable systems that were assigned slightly lower probabilities (low recall).
The appropriate threshold depends on the application (e.g., if predicting DNA matches for crime cases, one might set a high threshold as above to have confidence in predicted matches).

When comparing two models, one can plot pairs of precision and recall scores for all possible thresholds to generate a precision-recall curve.
A common scalar metric for comparing classifiers is the area under this curve (AUC), which would be unity for a perfect model.

\subsection{Algorithm Training} \label{training}
After experimenting with several machine learning algorithms (random forest and support-vector machine implementations in the {\tt Python scikit-learn} library), we found that gradient-boosted decision trees (GBDT\footnote{GBDT algorithms create and combine large numbers of individually weak but complementary classifiers to yield a robust estimator \citep{Friedman01}}) {\tt XGBoost v0.6} \citep{Chen16} yielded significantly higher AUC values for our dataset.

A recurring theme in machine learning is that of ``overfitting," an algorithm's tendency to latch onto irrelevant idiosyncrasies in the training set that cause it to predict poorly on unseen examples.
Different algorithms therefore try to penalize overly complicated models in an effort to retain only the broad features that are likely to generalize well.
In practice, the user navigates this balance between simplicity and complexity empirically, by tuning an algorithm's ``hyperparameters" that mediate this tradeoff, training it, and checking performance (Sec.\:\ref{metrics}) on unseen data; this process, together with trying different features to maximize performance, is known as cross-validation. 
In order to rule out the possibility of (sometimes subtle) mistakes in cross-validation yielding overly optimistic performance metrics, it can be good practice to assign a subset of the data to a holdout (test) set that is never seen by the algorithm during cross-validation.
Evaluation of the final model on the holdout set therefore provides robust metrics of the trained algorithm's expected performance on unseen examples; consistency between the cross-validation and test scores also suggests a robust cross-validation methodology.
In our case, we randomly assigned 1500 systems to a holdout test set, and used only the remaining 3500 for cross-validation.

A typical technique to reduce statistical fluctuations when comparing the performance of different sets of hyperparameters is $k$-fold cross validation.
One begins by splitting the training examples into $k$ evenly sized groups; then, for each group, one trains the model on the remaining $k-1$ chunks, and uses the remaining group to evaluate performance.
The scores from the $k$ folds are then averaged, reducing the variance in the estimate.
Finally, it is generally good practice to use stratified cross-validation, whereby one ensures that each of the $k$ folds is assigned an approximately equal number of samples from each class (stable and unstable).

{\tt XGBoost} has several hyperparameters, so we sequentially performed grid searches through 2-dimensional cuts through the parameter space, evaluating performance through the precision-recall AUC (Sec.\:\ref{metrics}) using stratified, 5-fold cross-validation on the training set of 3500 examples.
Values of the final adopted hyperparameters for the algorithm are discussed in Sec.\:\ref{IC} and \ref{shortintegrations} can be found in Table \ref{hyper}.

\begin{table}[h]
\begin{center}
\caption{Hyperparameters used for the initial-conditions (IC) model (Sec.\:\ref{IC}) and short-integrations (SI) model (Sec.\:\ref{shortintegrations}), and their associated performance. Scores for the Lissauer-family (LF) model shown for comparison. \label{hyper}}
\begin{tabular}{ c|c|c|c}
 	& LF & IC & SI \\
  \hline \hline			

  {\tt base\_score} & & 0.5 & 0.5\\
  {\tt colsample\_bylevel} & & 1 & 1\\
  {\tt colsample\_bytree} & & 1 & 1\\
  {\tt gamma} & & 0 & 0\\
  {\tt learning\_rate} & & 0.001 & 0.00359 \\
  {\tt max\_delta\_step} & & 0 & 0 \\
  {\tt max\_depth} & & 6 & 8 \\
  {\tt min\_child\_weight} & & 1.0 & 1.2 \\
  {\tt missing} & & None & None \\
  {\tt n\_estimators} & & 5000 & 5000 \\
  {\tt objective} & & bin:log & bin:log \\
  {\tt reg\_alpha} & & 0 & 0 \\
  {\tt reg\_lambda} & & 1 & 1 \\
  {\tt scale\_pos\_weight} & & 1 & 1\\
  {\tt seed} & & 27 & 27 \\
  {\tt subsample} & & 0.4 & 0.5\\
  \hline \hline
  AUC (Cross-validation) & & 0.84 $\pm$ 0.01 & 0.91 $\pm$ 0.01\\
  AUC(Test) & 0.77 & 0.84 & 0.90 \\
  Recall (90 \% Precision) & 0.30 & 0.52 & 0.68\\
    \hline  
\end{tabular}
\end{center}
\end{table}

\section{Results} \label{results}
\subsection{Model 1: Learning from Initial conditions} \label{IC}
We begin by considering as features the initial orbital elements and periods of each planet, as well as the interplanetary separations between adjacent planets in units of mutual Hill radii (23 features). 
We then trained an {\tt XGBoost} classifier on these features (Sec.\:\ref{training}), allowing us to predict the stability of each system in the test set in the form of an estimated probability. 

As discussed in Section~\ref{metrics}, a threshold probability is required for classifying a system as stable/unstable, and is a subjective choice that depends on the desired qualities of the classifier.
For our purposes we argue it is logical to adopt a conservative threshold, in the sense that if the model predicts stability, there is a strong likelihood that the system is actually stable (high precision).
This follows from the fact that it is computationally much faster to verify that a system is unstable (on short timescales) than it is to check that it is stable (on long timescales).
We choose to require a precision of $90\%$ on our test dataset, which corresponded to the model only classifying systems as stable if their predicted stability probability is larger than a 0.785 threshold \footnote{The reason this deviates from the desired precision is that the probability that the model outputs is heuristic, and is not necessarily an accurate measure of the likelihood that the system is actually stable.}.

\begin{figure}
 \centering \resizebox{0.99\columnwidth}{!}{\includegraphics{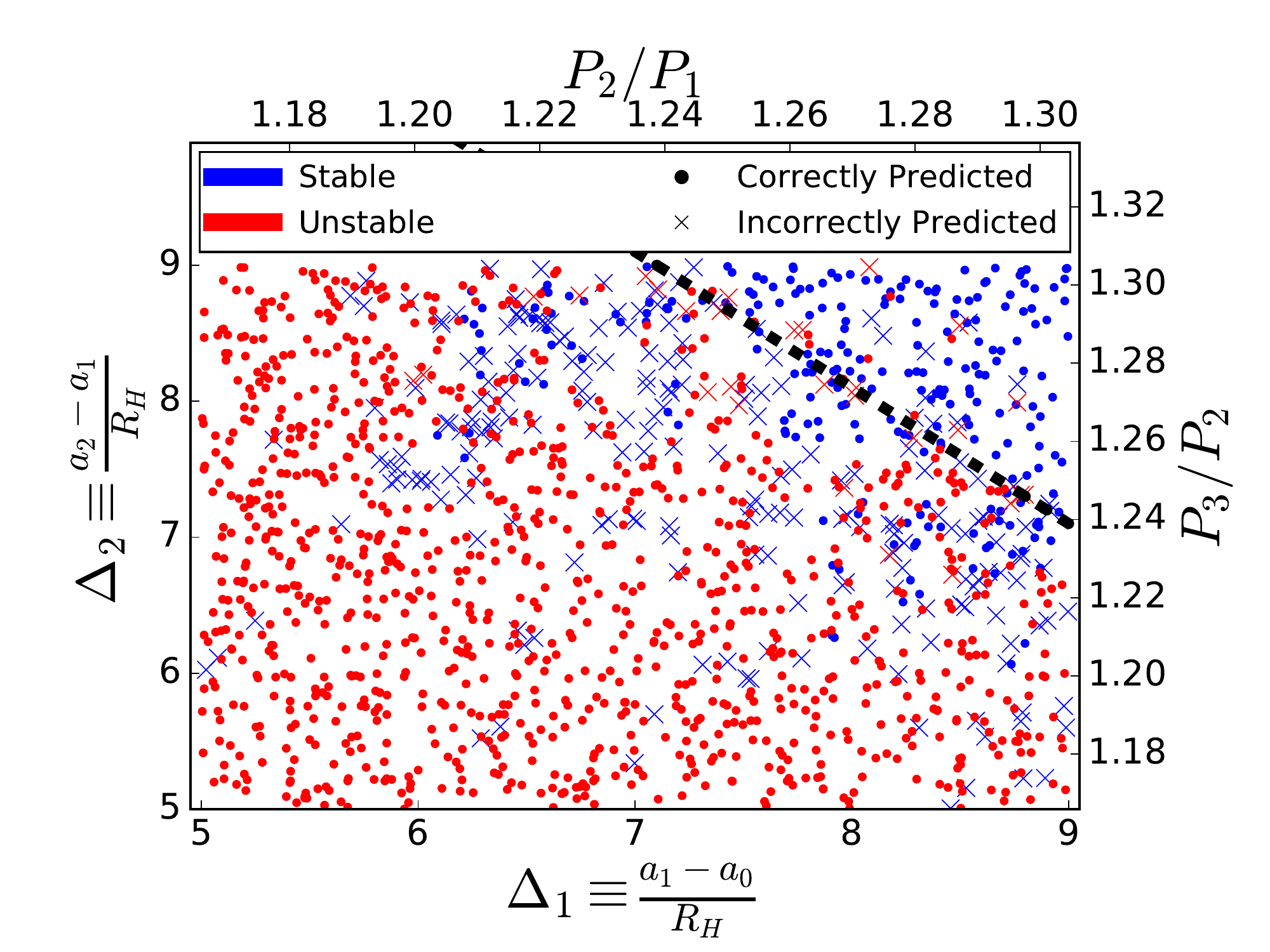}}
 \caption{
     Performance on the test dataset using a model trained on systems' initial conditions.
     Stable systems are marked blue, unstable systems are marked red.  
     Correctly classified systems are plotted as circles, incorrect predictions are marked as crosses.
     The bottom and left axes show Hill-sphere separations $\Delta$ for the inner and outer planet pairs, respectively.
     The top and right axes correspond to period ratios between the planet pairs.
     The dashed black line corresponds to the Lissauer-family model $\Delta_1 + \Delta_2 > 16.1$.    
    \label{ariplot}}
\end{figure}

Since previous works have identified the Hill separations between adjacent planets as important features \citep{Chambers96, Marzari14}, we plot the performance of the model projected onto this 2D plane (Fig.\:\ref{ariplot}).

Looking at the dashed, black line in Figure~\ref{ariplot}, one can see that to first order, the model's prediction boundary roughly obeys the relation $\Delta_1 + \Delta_2 > 16.1$.
This is in fact the form of the simple criterion suggested by \cite{Lissauer11}, who quote $\Delta_1 + \Delta_2 > 18$ for stability on timescales of $10^9$ orbits. 
Because we consider stability on shorter timescales, the threshold number of Hill radii should be adjusted for a fair comparison.
We term models of the form $\Delta_1 + \Delta_2 > x$ ``Lissauer-family" models, and find 16.1 is the threshold for Lissauer-family models that yields a precision of 90\% on this dataset.

As stated above, a conservative probability threshold has the disadvantage that the model will misclassify many stable systems as unstable (low recall).  
This is easily seen by considering different Lissauer-family models (dashed black line in Fig.\:\ref{ariplot}), i.e. imposing different threshold values than 16.1 and generating lines parallel to the one plotted.
Larger threshold values ensure that a larger fraction of the systems to the right of the boundary (i.e., those predicted stable) are in fact stable (blue), leading to higher precision.
However, this lowers the recall, since now fewer of the stable systems (blue) are predicted to be stable by the model (i.e., lie to the right of the line).
For a fixed precision of 90\%, the machine learning model has a significantly higher recall ($52\%$) than the Lissauer-family model ($30\%$). 
This is because the machine learning model can use information in the features not visible in this 2D projection to make better predictions.
The uncertainty in the recall should be comparable to the rms variations in the AUC calculated across the k folds during cross-validation, or $\approx 1\%$ (Table \ref{hyper}).

\subsection{Model 2: Generating Features from Short Integrations} \label{shortintegrations}

An important factor determining the performance of a machine learning algorithm is the quality of the features it is provided for each of the training examples.
To this end, we improved upon the previous model (Section~\ref{IC}) by generating new features from short N-body integrations.
To create the new features, we performed simulations over $10^4$ orbits (0.1\% of the simulation timescale probed) for each of the 5000 systems in the dataset, and recorded each planet's orbital elements and Lyapunov time every 5 orbits.

Because we suspect that the instability is driven by overlapping mean-motion resonances, we first generated features that capture semimajor-axis variations, which would approximately vanish if the dynamics were purely secular \citep{Murray99}.  
In particular, we generated features for the standard deviation and maximum value of each planet's semimajor axis over the $10^4$ orbits, normalized to the mean value over the same period  ({\tt std\_ai} and {\tt max\_ai}, where {\tt i} denotes the planet number).  
We also generated features for these quantities over only the first 50 orbital periods to capture variations on orbital timescales ({\tt std\_window\_ai} and {\tt max\_window\_ai}).
Furthermore, we capture any drifts by generating slope features from linear fits to each of the three planets' semimajor axes, normalized to the mean semimajor axis divided by the integration time ({\tt slope\_ai}).
For the eccentricities, we took the mean, standard deviation, maximum and minimum values over the full $10^4$ orbits, normalized to the eccentricity the planet would require to reach its nearest neighbor's semimajor axis ({\tt avg\_ei}, {\tt std\_ei}, {\tt max\_ei}, {\tt min\_ei}).
For the Lyapunov time we generated a single feature corresponding to the value measured at the end of the integration, normalized to the innermost orbital period ({\tt LyapTime}).
Finally, we added features for the two pairs of initial Hill-radius separations ({\tt daOverRH1}, {\tt daOverRH2}), and for the minimum and maximum initial Hill-radius separations ({\tt mindaOverRH}, {\tt maxdaOverRH}).
We experimented with features describing orbital inclinations, but they did not significantly improve the models.

A summary of the features can be found in Table \ref{features}, ordered by their importance.
We quantify the importances through the {\tt Gain} value recorded by {\tt XGBoost}, which corresponds to the gain in accuracy that a given feature provides when it is introduced into the underlying decision trees used by the algorithm.  
The units are normalized so that the gains sum to 100.
We find that the variations in the middle planet's semimajor axis are the most informative in this sense.
This suggests that the instabilities in these closely packed systems are driven by the overlap of mean motion resonances (which change the semimajor axes), rather than secular effects (which would keep the semimajor axes constant).

\begin{table}[h]
\begin{center}
\caption{`Short-Integration' Model Feature Importances. See text for a description of the gain and of the features. \label{features}}
\begin{tabular}{ c|c||c|c}
Feature & Gain & Feature & Gain \\
  \hline \hline			
  {\tt max\_a2} & 20.3 & {\tt max\_window\_a3} & 1.6 \\
  {\tt std\_a2} & 8.3 & {\tt std\_window\_a3} & 1.6 \\
  {\tt mindaOverRH} & 2.6 & {\tt std\_e1} & 1.5 \\
  {\tt maxdaOverRH} & 2.6 & {\tt std\_e2} & 1.5 \\
  {\tt std\_a3} & 2.3 & {\tt std\_window\_a1} & 1.5 \\
  {\tt max\_e2} & 2.3 & {\tt max\_e1} & 1.5 \\
  {\tt daOverRH1} & 2.3 & {\tt slope\_a3} & 1.5 \\
  {\tt daOverRH2} & 2.2 & {\tt slope\_a1} & 1.5 \\
  {\tt max\_e3} & 2.1 & {\tt max\_window\_a2} & 1.5 \\
  {\tt std\_a1} & 1.9 & {\tt std\_e3} & 1.4 \\
  {\tt max\_a3} & 1.9 & {\tt min\_e3} & 1.4 \\
  {\tt avg\_e2} & 1.8 & {\tt avg\_e1} & 1.4 \\
  {\tt avg\_e3} & 1.8 & {\tt max\_window\_a1} & 1.4 \\
  {\tt max\_a1} & 1.8 & {\tt min\_e2} & 1.4 \\
  {\tt LyapTime} & 1.7 & {\tt min\_e1} & 1.3 \\
  {\tt std\_window\_a2} & 1.6 & {\tt slope\_a2} & 1.3 \\
  \hline  
\end{tabular}
\end{center}
\end{table}

\begin{figure}
 \centering \resizebox{0.99\columnwidth}{!}{\includegraphics{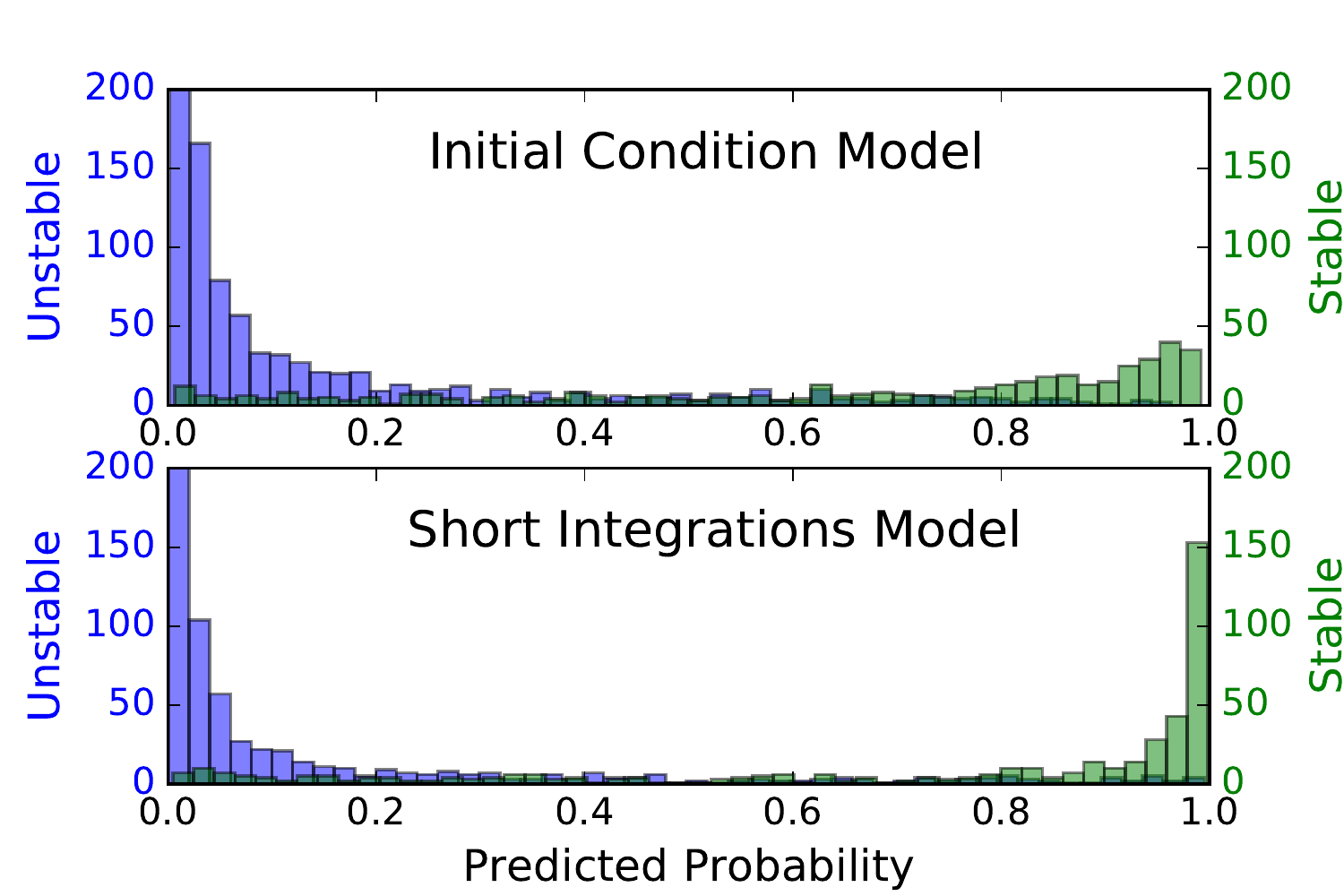}}
 \caption{
     Comparison of predictions of the initial-condition and short-integration models on the test set.
     Stable systems are shown in green, unstable systems are shown in blue, and the model-predicted probability of stability for each system is shown along the x-axis.
     The leftmost blue bin is cut off to render smaller bins visible--in the top panel it reaches 395, and in the bottom panel 640.
    \label{dianaplot}}
\end{figure}

Finally, we compare the performance of this `short-integration' model to the previous `initial-condition' model.
Fig.\:\ref{dianaplot} shows that while both models often assign unstable systems (blue bins) a low predicted probability, the `initial-condition' model assigns stable systems (green bins) a wide range of predicted probabilities, translating to a lower recall.
In contrast, the `short-integration' model more confidently assigns high predicted probabilities to stable systems, better separating the two classes.
Again setting the predicted probability threshold so as to obtain 90\% precision, the recall improves to 68\%.

In summary, generating improved features from short integrations and adopting {\tt XGBoost} as our algorithm provided our largest AUC gains (Table \ref{hyper}).

\section{Discussion \& Conclusion} \label{conclusion}
In this investigation, we numerically integrated a dataset of 5000 three-planet systems over $10^7$ orbits.
We then trained machine learning algorithms to classify systems' orbital stability on this timescale.
In particular, we trained two models using the {\tt XGBoost} algorithm: an `initial-conditions' model (Sec.\:\ref{IC}) that learned only from the system's initial orbital elements, and a `short-integration' model (Sec.\:\ref{shortintegrations}) that generated features from short N-body integrations.
We then compared their performance to `Lissauer-family models' that require the sum of the interplanetary separations (expressed in mutual Hill radii) to be greater than a particular threshold.

\begin{figure}
 \centering \resizebox{0.99\columnwidth}{!}{\includegraphics{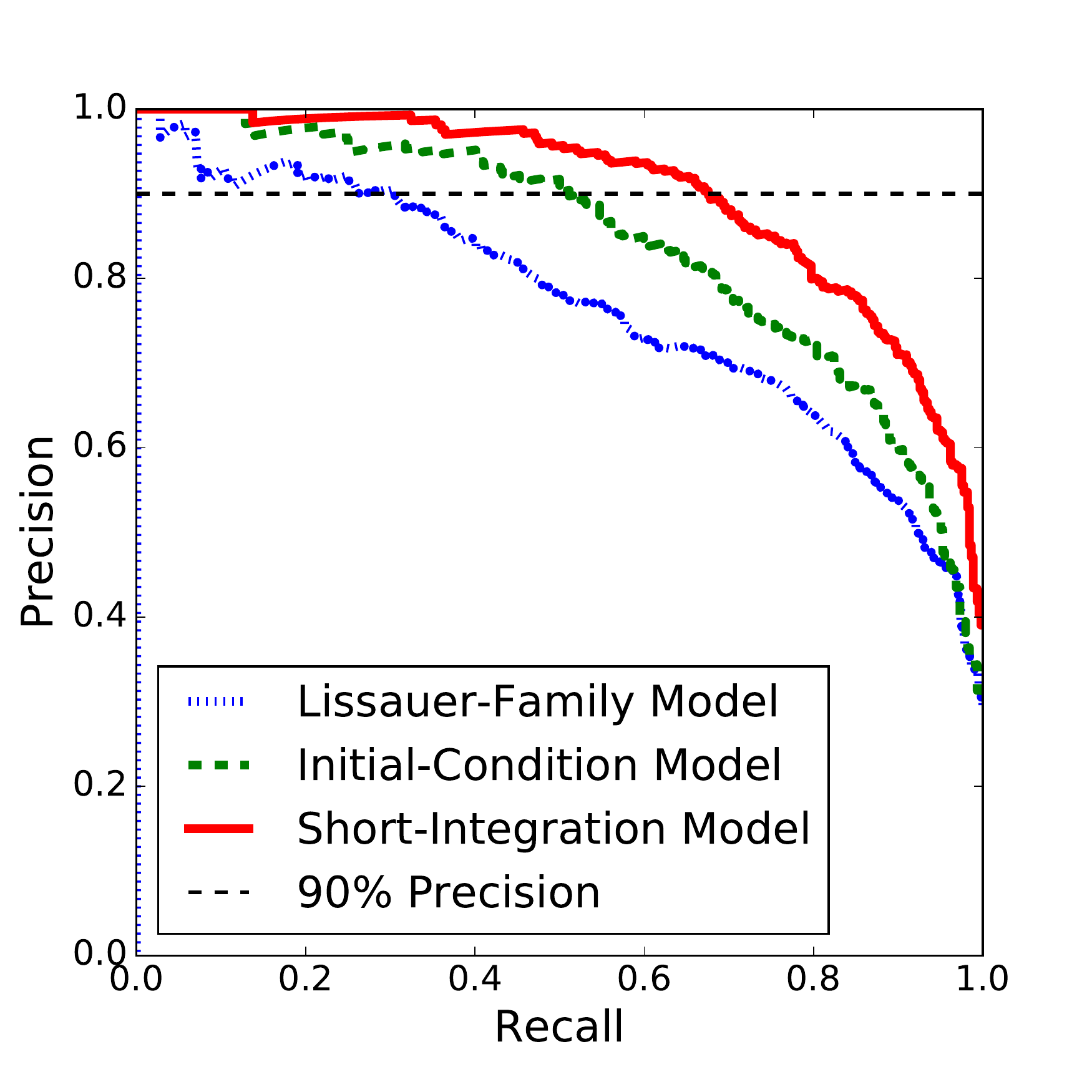}}
 \caption{
     Precision-recall curves for each model in this paper, generated from all possible values of the model's threshold classification probability.
     The Lissauer-family models predict stability if the sum of the Hill-sphere separations is greater than a particular threshold, and the corresponding curve was generated by considering all possible threshold values.
     The horizontal black-dashed line is the 90\% precision requirement we imposed on all models in this paper. 
    \label{prcurves}}
\end{figure}

We summarize the investigated models' performances in Fig.\:\ref{prcurves}, which plot values for the respective classifier's precision and recall for all possible values of the probability threshold above which the model labels a system as stable.
As discussed above, the appropriate choice of this threshold depends on the desired qualities of the classifier.
Above we advocated for conservative models that are correct 90\% of the time when a system is predicted stable (90\% precision). 

Our best machine-learning model (right column of Table\:\ref{hyper}):
\begin{itemize}
	\item Dominates other models at all threshold values.
	\item Recovers 2.25 as many of the stable systems in the test set (has 2.25 times higher recall) as a Lissauer-family model at a fixed precision of 90\%.
	\item Is three orders of magnitude faster than direct N-body integration. Performing the short integration, generating the features and evaluating the model for a given system takes $\sim$ 1 second with current technology.
\end{itemize}

We note that while we have focused on making binary predictions (stable vs. unstable) for straightforward model comparisons, the XGBoost classifier can instead output a predicted confidence for a given system between 0 and 1.
This has more information and could for example be used to estimate a prior probability distribution for fitting orbital data.

An important limitation of this work are the fixed masses (5 Earth masses around a solar-mass star) and the comparatively short integration timescales ($10^7$ orbits).
Our results strongly motivate investing computational time to generate datasets over longer timescales with a range in masses.

Such classifiers could be used in many of the ways that direct N-body integrations currently are employed, but these models' dramatically improved efficiency would allow much faster and complete explorations of parameter space, e.g.:
\begin{itemize}
	\item Mapping out the stability boundary in mass-eccentricity space for observed transiting systems.
	\item Mapping out the parameter space that unseen planets could stably inhabit in a given system to guide observational follow-up strategies or theoretical considerations.
	\item Vetting low signal-to-noise detections through stability constraints.
	\item Generating a prior probability distribution describing the allowable regions of phase space for statistical or theoretical investigations.
	\item As a stopping condition for simulations once they achieve a dynamically long-lived configuration.
\end{itemize}

Such tools may be of particular interest for the upcoming Transiting Exoplanet Survey Satellite (TESS).
While transit-timing variations (TTVs) have been powerful tools for constraining the masses and eccentricities of near-resonant Kepler systems \citep[e.g.,][]{Ford12, Steffen13, Deck15}, TESS' shorter time baselines and planets' smaller semimajor axes will likely render such analyses difficult or impossible.
It is therefore likely that in many systems, stability considerations will provide the strongest constraints on planetary masses and eccentricities, and this will be important for guiding the substantial radial-velocity follow-up efforts from the ground. 

More broadly, we have shown that machine learning can be a powerful tool for high-dimensional classification problems in dynamics.
But in addition to their predictive power, our models also revealed new insights into the underlying physics.
In particular, the most informative features in our model based on short integrations were the variations in the middle planet's semimajor axis.
This suggests that the orbital instabilities are driven by the overlap of mean motion resonances (which vary the semimajor axes) rather than the secular chaos at work in our own solar system \citep{Lithwick11secularchaos, Batygin15}, which keeps semimajor axes approximately constant.

\section{Acknowledgements}
We would like to thank David Kipping for an insightful review and Fred Rasio for helpful comments. Both greatly improved the accuracy and presentation of this work. D.T., M.A. and C.H. were supported by fellowships through the University of Toronto's Centre for Planetary Sciences, and D.T. also gratefully acknowledges support from the Jeffrey L. Bishop Fellowship.

\end{document}